\documentclass[twocolumn,showpacs,showkeys,preprintnumbers,amsmath,amssymb,refcheck]{revtex4}
\usepackage{graphicx}
\usepackage{dcolumn}
\usepackage{bm}

\begin{document}

\preprint{ }
\title{
Hamiltonian of Homonucleus Molecules for NMR Quantum Computing}

\author{Yasushi Kondo$^{1}$, Mikio Nakahara$^{1}$, Kazuya Hata$^{1}$,
and Shogo Tanimura$^{2}$\\}

\affiliation{%
$^{1}$Department of Physics, Kinki University, 
Higashi-Osaka 577-8502, Japan\\
$^2$Graduate School of Engineering, Osaka City University, 
Sumiyoshi-ku, Osaka 558-8585, Japan\\
}%

\date{\today}

\begin{abstract}

We derive the Hamiltonian in the rotating frame for NMR quantum computing 
with homonucleus molecules as its computational resource. The Hamiltonian thus
obtained is different from conventional
Hamiltonians that appear in literature. It is shown that
control pulses designed for heteronucleus spins can be 
translated to pulses for homonucleus spins
by simply replacing hard pulses by soft pulses with properly chosen pulse
width. To demonstrate the validity of our Hamiltonian, we conduct several
experiments employing cytosine as a homonucleus molecule.
All the experimental results 
indicate that our Hamiltonian accurately describes the dynamics of the
spins and that  the conventional Hamiltonian fails.
Finally we use our Hamiltonian for
precise control of field inhomogeneity compensation
with a pair of $\pi$-pulses.

\end{abstract}



\maketitle

\section{Introduction}

Quantum computation currently attracts a lot of attention
since it is expected to solve some of computationally
hard problems for a conventional digital computer~\cite{ref:1}. 
Numerous realizations of a quantum computer have been proposed to date.
Among others, a liquid-state NMR 
(nuclear magnetic resonance)
quantum computer is regarded as
most successful. Early experiments demonstrated 
         quantum teleportation~\cite{nmr1}, 
         quantum search algorithm~\cite{nmr2},
         quantum error correction~\cite{nmr3}, and
         simulation of a quantum mechanical system~\cite{nmr4}. 
Undoubtedly, demonstration of Shor's factorization algorithm~\cite{VSB01}
is one of the most remarkable achievements in NMR quantum computation. 
Although the number of admissible qubits in a liquid-state NMR quantum 
computer is suspected to be limited up to about ten due to poor 
spin polarization at a room temperature, 
a liquid-state NMR quantum computer
is one of few quantum computers that are capable of running
nontrivial quantum algorithms thanks to well established NMR technology.

Since the number of qubits within heteronucleus spins
is practically limited to two or three,
the use of homonucleus spins is inevitable
if we try to equip an NMR with a large number of qubits.
It should be pointed out, however, that
liquid-state NMR of homonucleus molecules
is still poorly understood and literature dealing with this subject often
lacks solid ground. 
Although the product operator formalism \cite{ernst} has been 
extensively employed to implement quantum algorithms with
an NMR quantum computer,
people overlooked significance of the genuine Hamiltonian.
Actually there is a subtle difference between 
the conventionally used Hamiltonian 
and the proper Hamiltonian for homonucleus spins.
It is, therefore, urgently required to establish
theoretical foundation underlying
a liquid-state NMR quantum computer with homonucleus molecules.

Suppose we would like to implement a quantum algorithm whose unitary
matrix representation is $U_{\rm alg}$. 
If the Hamiltonian $ H $ depends on the control parameters, which
we write collectively as $ \gamma(t) $,
the time evolution operator is given by
\begin{equation}
        U[\gamma(t)] = {\mathcal T} \exp\left[-i
        \int_0^T H(\gamma(t)) dt\right],
\end{equation}
where ${\mathcal T}$ stands for the time-ordering product.
We use the natural unit in which $\hbar = 1 $. 
Optimal control of the quantum computer requires 
a control function $ \gamma(t) $
that produces the specified quantum algorithm 
$ U[\gamma(t)] = U_{\rm alg} $ 
in the shortest possible time $T$. 
Recently, numerical scheme to find the optimal control 
has been worked out for fictitious Josephson junction qubits, 
where polygonal paths in the parameter space has been
utilized~\cite{qaa4,qaa6}. 
For time-optimal control of an NMR quantum computer,
another method employing the Cartan decomposition of SU($ 2^n $)
has been proposed~\cite{ref:kg}
and has been demonstrated
experimentally with a two-qubit heteronucleus molecule~\cite{qaa}.
We note that exact optimal control has been found
for holonomic quantum computation
in an idealized situation~\cite{qaa3.5}.

This paper has three aims:  
(1) to provide the theoretical foundation for an NMR quantum computer
with homonucleus molecules,
(2) to show that any pulse sequence designed for heteronucleus molecules 
 can be translated into that for homonucleus molecules, and
(3) to demonstrate experimentally that our Hamiltonian accurately describes the
dynamics of the spins.
For these purposes, we carefully examine the 
Hamiltonians for NMR spin dynamics.
Although the Hamiltonian for a homonucleus molecule
is the same as the one for a heteronucleus molecule in the laboratory frame,
the former looks quite different from the latter in a rotating frame.

This paper is organized as follows.
In section II
we study Hamiltonians of homonucleus as well as heteronucleus molecules.
We carefully examine how they are transformed in a rotating frame
and what is appropriate approximation to be employed. Surprisingly,
our resulting Hamiltonian is different from the conventional Hamiltonian.
In section III, we conduct several experiments to verify our analysis
by taking cytosine as an example of homonucleus molecules.
We execute the Deutsch-Jozsa algorithm, execute the pulse sequence
for pseudo-pure state preparation, and verify the robustness of two-qubit
entangling operations. As an application of the correct form of the
Hamiltonian we implement field inhomogeneity 
compensation using a pair of $\pi$-pulses in section IV. 
Section V is devoted to conclusions and discussion. 

\section{Hamiltonian in Rotating Frame}

In this section, we write down 
the Hamiltonian of spin dynamics in the laboratory frame 
and transform it to the one in a rotating frame.
Although 
the Hamiltonian for a homonucleus molecule
has the same form as the one for a heteronucleus molecule
in the laboratory frame,
Hamiltonians in a rotating frame differ from each other.

We restrict ourselves within two-qubit molecules for simplicity.
Generalization to molecules with more qubits is straightforward. 
As an example of heteronucleus molecules,
we refer to $^{13}$C-labeled chloroform.
The qubits are spins of $^{13}$C and H nuclei.
We take cytosine solved in D$_2$O as an example of homonucleus molecule.
The qubits are spins of two hydrogen nuclei (protons) in this case. 

\subsection{Heteronucleus molecule}

\subsubsection{Experimental setup}

A liquid-state NMR
consists of three parts as described in \cite{NMR_textbook}. 
The first part is magnetic coils;
a superconducting coil to generate a homogeneous static magnetic field
and a normal conducting coil 
to generate temporally controlled field gradients. 
The second part contains resonance circuits 
for applying radio frequency (rf) magnetic fields to the sample. 
They are also used to pick up rf signals from the sample. 
The third part is an assembly of electronic circuits to 
feed rf pulses into the resonance circuits
and to detect the signals picked up by the coils.

\begin{figure}[tb]
\includegraphics[bb=0 0 470 630,width=8cm]{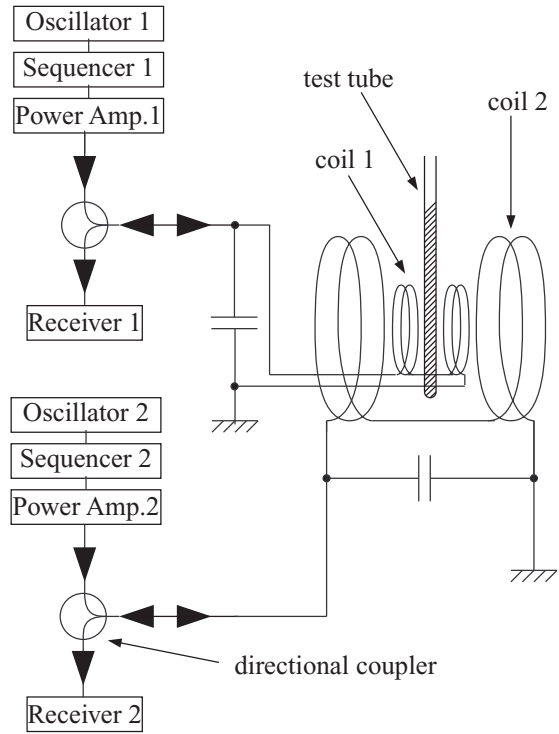}
\caption{
NMR setup for heteronucleus molecules.
\label{hetero}
}
\end{figure}

The NMR setup for heteronucleus molecules
is shown schematically in Fig.~\ref{hetero}.
The first and second spins have 
respective resonance frequencies 
$ \omega_{0,i} $ ($ i=1,2 $),
which are also called Larmor frequencies.
Their resonance frequencies are widely different for heteronucleus
molecules under consideration.
Hence two sets of resonance circuits and assembly of electronic circuits 
are required.
The large difference of the resonance frequencies, 
$ \Delta \omega_0 = \omega_{0,2} - \omega_{0,1} $, 
allows us to address each spin individually with a short pulse.

The oscillator $i$ $(i=1,2) $
in the third part generates an rf electric wave with frequency 
$ \omega_{{\rm rf},i} $.
The sequencer $i$ modulates the rf wave 
to shape a designed pulse.
A typical temporal duration of a pulse, 
which is called the pulse width, 
is of the order of 10~$\mu$s.
The rf pulses are amplified and fed into the resonance coil $i$, 
which generates rf magnetic fields
applied to the sample in the test tube. 
Precession of spins in molecules appears as
rotation of magnetization of the sample
and induces a signal at the coil $i$.
The receiver $i$ detects the signal.
The directional coupler
prevents transmission of the rf pulse from 
the amplifier to the  receiver.

\subsubsection{Heteronucleus molecule in rotating frame}
\label{h_hetero_rf}

The two-qubit Hamiltonian in the laboratory frame is
\begin{equation}
\label{H in lab}
H = H_0 + H_{{\rm rf},1} + H_{{\rm rf},2}.
\end{equation}
Here the system Hamiltonian $H_0$ is defined as~\cite{rmp_chuang}
\begin{equation}
H_0 =
- \omega_{0,1} I_z \otimes I 
        - \omega_{0,2} I \otimes I_z 
+ \!\! \sum_{k=x,y,z} \!\! J I_k \otimes I_k,
\label{system H}
\end{equation}
where $I_k = \sigma_k/2$, 
$\sigma_k$ being the $k$-th Pauli matrix, and 
$I$ is the unit matrix of dimension two.
The first two terms in $ H_0 $ describe free precession of the spins
in a static magnetic field
while the third term describes the intramolecule spin interaction 
with coupling strength $ J $. 

$ H_{{\rm rf}, i} $ ($i=1,2$) represents the action
of the rf magnetic field generated by the coil $ i $
and hence is called the control Hamiltonian.
Their explicit forms are
\begin{eqnarray}
H_{{\rm rf},1}
\!\! &=& \!\!
- 2 \omega_{1,1} \cos( \omega_{{\rm rf},1} t - \phi_1)
        ( I_x \otimes I + g I \otimes I_x),       
\label{h_rf_lab1}  
\\
H_{{\rm rf},2}
\!\! &=& \!\!
- 2 \omega_{1,2} \cos( \omega_{{\rm rf},2} t - \phi_2)
        ( g^{-1} I_x \otimes I + I \otimes I_x).
        \;\;
\label{h_rf_lab2}
\end{eqnarray}
Here,
the amplitude of the rf pulse $ \omega_{1,i} $, 
the frequency of the pulse $ \omega_{{\rm rf},i} $ 
and the phase of the pulse $\phi_i$ 
are controllable parameters. 
We may assume, without loss of generality, that
the rf field is applied along the $x$-axis in the laboratory frame.
In the above equations 
we introduced the ratio of resonance frequencies of the two nuclei,
\begin{equation}
g = \frac{\omega_{0,2}}{\omega_{0,1}}.
\label{g}
\end{equation}

We shall examine the transformation law of the Hamiltonians
from the laboratory frame to a rotating frame. 
The spin dynamics in the laboratory frame 
is governed by the Liouville equation
\begin{equation}
i \frac{d \rho}{dt} = [ H, \rho ],
\end{equation}  
where $\rho$ is the density matrix of the system under
consideration.
The unitary operator
\begin{equation}
\label{rot}
U = 
\exp ( -i \omega_{{\rm rot},1} I_z t) \otimes 
\exp ( -i \omega_{{\rm rot},2} I_z t)
\end{equation}
transforms $ \rho $ into 
the density matrix $ \tilde{\rho} $ in the rotating frame as
\begin{equation}
\tilde{\rho} =  U \rho \, U^{\dagger}.
\end{equation}
Note that we can choose the rotation angular velocities
$ \omega_{{\rm rot},i} $ ($i=1,2$) arbitrarily.
The time evolution of the system is now governed by
\begin{eqnarray}
i  \frac{ d \tilde{\rho} }{dt} = [ \tilde{H}, \tilde{\rho} ]
\end{eqnarray}
with the transformed Hamiltonian
\begin{eqnarray}
\tilde{H}
&=& 
U H U^{\dagger}
- i U \frac{d}{dt}U^{\dagger}
= \tilde{H}_0
+ \tilde{H}_{{\rm rf},1}
+ \tilde{H}_{{\rm rf},2}.
\end{eqnarray}
Here the transformed system Hamiltonian is
\begin{eqnarray}
\label{h0_r}
\tilde{H}_0 
&=& 
U H_0 U^{\dagger} - iU \frac{d}{dt}U^{\dagger} 
\nonumber\\
&=&
- ( \omega_{0,1} - \omega_{{\rm rot},1} ) I_z \otimes I   
\nonumber \\ &&
- ( \omega_{0,2} - \omega_{{\rm rot},2} ) I   \otimes I_z 
+ J I_z  \otimes I_z 
\nonumber \\
&& + \left( 
        \begin{array}{cccc}
             0 & 0 & 0 & 0 \\
             0 & 0 & \frac{J}{2}e^{i \Delta \omega_{\rm rot} t} & 0 \\
             0 & \frac{J}{2}e^{-i \Delta \omega_{\rm rot} t} & 0 & 0 \\
             0 & 0 & 0 & 0 
        \end{array}
        \right),
\end{eqnarray}
where 
$ \Delta \omega_{\rm rot} 
\equiv \omega_{{\rm rot},2} - \omega_{{\rm rot},1} $.
The transformed control Hamiltonians 
$ \tilde{H}_{{\rm rf},i} $ will be given later. 
If we take the frame co-moving with each spin, which has
the angular velocities $ \omega_{{\rm rot},i} = \omega_{0,i} $,
the first two terms in Eq.~(\ref{h0_r}) vanish.   
In the case of heteronucleus molecules, 
the condition $ \left| \Delta \omega_0 \right| \gg J $ 
is always satisfied 
and thus the matrix elements in the last line also vanish
after averaging over time. For example, 
$ \left| \Delta \omega_0 \right|/2 \pi \sim 400$~MHz while 
$ J /2 \pi\sim 200$~Hz for 
$^{13}$C-labeled chloroform at 11~T, for which 
$\left| \Delta \omega_0 \right| / J \sim 10^6$.
Therefore $ \tilde{H}_0 $ is well approximated by
\begin{equation}
\label{h0_r2}
\tilde{H}_0 = J I_z \otimes I_z.
\end{equation}

When the resonance and co-rotating conditions
$ \omega_{{\rm rf},i} = \omega_{0,i} = \omega_{{\rm rot},i} $
are satisfied,
the control Hamiltonians in the rotating frame
\begin{equation}
\tilde{H}_{{\rm rf},i} =
U H_{{\rm rf},i} U^{\dagger} 
\end{equation}
are approximately given as
\begin{eqnarray}
\tilde{H}_{{\rm rf},1} &=& 
- \omega_{1,1} 
( \cos \phi_1 \, I_x \otimes I 
+ \sin \phi_1 \, I_y \otimes I ),
\label{h_rf_r1}
\\
\tilde{H}_{{\rm rf},2} &=& 
- \omega_{1,2} 
( \cos \phi_2 \, I \otimes I_x
+ \sin \phi_2 \, I \otimes I_y ) 
\label{h_rf_r2}
\end{eqnarray}
after dropping terms rapidly oscillating with 
frequencies $ 2 \omega_{0,i} $ and $ \Delta \omega_0 $.
Note that the factor 2 in front of $ \omega_{1,i} $ 
in Eqs.~(\ref{h_rf_lab1}) and (\ref{h_rf_lab2})
has disappeared
in Eqs.~(\ref{h_rf_r1}) and (\ref{h_rf_r2}).
This is physically understood as discussed in \cite{NMR_textbook};
a linearly polarized rf magnetic field oscillating with frequency $ \omega_{\rm rf} $
is a superposition of two circularly polarized fields 
with frequencies $ \pm \omega_{\rm rf} $ 
and the effect of the component with $- \omega_{\rm rf}$ is averaged to vanish.
It is also important to notice that a pulse with frequency 
$ \omega_{{\rm rf},i} $ influences only the spin $i$ 
and does not affect the other spin in the rotating frame. 
This is because $ \left| \Delta \omega_0 \right |$ 
is much larger than the inverse of the typical pulse width 
$ \sim 1/(10 \, \mu s) \sim 100 $~kHz and 
hence the rf pulse resonating with one spin
does not have spectral component which affects the other spin. 

In conclusion, the Hamiltonian for a heteronucleus molecule
in resonant magnetic fields 
is 
\begin{eqnarray}
\label{hetero_ham}
\tilde{H} 
&=& J I_z  \otimes I_z
        \nonumber\\
& & - \omega_{1,1}  
( \cos \phi_1 \, I_x \otimes I + \sin \phi_1 \, I_y \otimes I )
\nonumber\\ 
& & - \omega_{1,2}
( \cos \phi_2 \, I \otimes I_x + \sin \phi_2 \, I \otimes I_y )
\end{eqnarray}
in the rotating frame that has 
the angular velocities $ \omega_{{\rm rot},i} = \omega_{0,i} $.

\subsection{Homonucleus molecules}

\subsubsection{Experimental setup}

\begin{figure}[tb]
\includegraphics[bb=0 270 470 660,width=8cm]{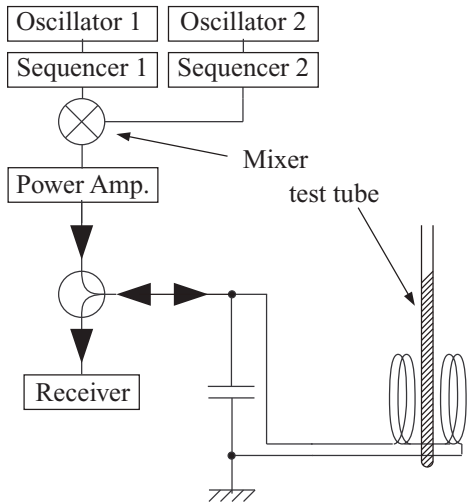}
\caption{
NMR setup for homonucleus molecules.
\label{homo2}
}
\end{figure}
The NMR setup for homonucleus molecules
is shown schematically in Fig.~\ref{homo2}. 
Because 
the difference of the resonance frequencies
$ \Delta \omega_0 =  \omega_{0,2} - \omega_{0,1}  $ 
is not large compared to $\omega_{0, i}$ in this case, 
a common resonance circuit and a power amplifier can be used to 
control both spins. For cytosine in D$_2$O, for
example, we find 
$ \left| \Delta \omega_0 \right| / 2 \pi \sim 765.0$~Hz 
while  $ \omega_{0,i} /2 \pi \sim 500$~MHz. 
Although the difference $ \Delta \omega_0 $ is small, 
it still allows us to address respective spins individually
provided that the pulse width is sufficiently long.

The oscillator $ i $ generates a continuous rf electric wave with
frequency $ \omega_{{\rm rf},i} = \omega_{0,i} $.
The sequencer shapes the continuous wave into pulses.
When addressing the two spins simultaneously,
a typical pulse width is of the order of 10~$\mu$s.
On the other hand, when addressing them individually,
a typical pulse width is of the order of 
$ 2 \pi/ \left| \Delta \omega_0 \right| \sim 1$~ms.
The rf pulses from the two sequencers are mixed and amplified. 
The coil generates magnetic fields and picks up signals from the sample
and the receiver detects the signals.
Due to close resonance frequencies $\omega_{0, i}$,
only one set of resonance circuit and receiver is necessary
for homonucleus molecules.

\subsubsection{Homonucleus molecule in rotating frame}

The Hamiltonian for homonucleus molecule in the laboratory frame
has the identical form to
the Hamiltonian for a heteronucleus molecule (\ref{H in lab}).
Even for homonucleus molecule
the condition $ \left| \Delta \omega_0 \right| \gg J $ 
is satisfied in general.
For example, in the case of cytosine in D$_2$O, 
$ \left| \Delta \omega_0 \right| / 2 \pi \sim 765.0$~Hz
while $ J /2 \pi \sim 7.1$~Hz, and thus 
the above condition is satisfied. 
Therefore, the approximation 
used in the derivation 
of the system Hamiltonian (\ref{h0_r2}) for a heteronucleus molecule
is also applicable 
to derivation of that for a homonucleus molecule.
Thus the system Hamiltonian of a homonucleus molecule
takes the form
\begin{equation}
\label{h0_homo}
\tilde{H}_0 = J I_z \otimes I_z.
\end{equation}
in the co-rotating frame of each spin.

The control Hamiltonian $ \tilde{H}_{{\rm rf},i} $ 
describes the action of the resonant magnetic field 
in the frame rotating with angular velocity 
$ \omega_{{\rm rot},i} = \omega_{0,i} = \omega_{{\rm rf},i} $.
Corresponding Hamiltonian
$ \tilde{H}_{{\rm rf},i} $ for homonucleus molecule
is considerably more complicated 
even when terms rapidly oscillating with frequencies
$ 2 \omega_{0,i} $ 
and $ \omega_{0,1} + \omega_{0,2} $ are averaged out as
\begin{eqnarray}
\tilde{H}_{{\rm rf},1}
&=& - \omega_{1,1} 
\Big[
  \cos \phi_1 \, I_x \otimes  I 
+ \sin \phi_1 \, I_y \otimes  I 
\nonumber \\
&& + g 
  \cos ( \Delta \omega_0 t + \phi_1) I \otimes I_x 
\nonumber \\
&& + g 
  \sin ( \Delta \omega_0 t + \phi_1) I \otimes I_y
\Big],
\label{h_rf_r_homo1}
\\
\tilde{H}_{{\rm rf},2}
&=& - \omega_{1,2}
\Big[
  \cos \phi_2 \, I \otimes I_x 
+ \sin \phi_2 \, I \otimes I_y )
\nonumber \\
&& + g^{-1} 
  \cos ( -\Delta \omega_0 t + \phi_2) I_x\otimes  I 
\nonumber \\
&& + g^{-1} 
  \sin ( -\Delta \omega_0 t + \phi_2) I_y \otimes I
\Big].
\label{h_rf_r_homo2}
\end{eqnarray}

If we further assume that the pulse width $\tau$ are 
long enough so that even slowly oscillating terms 
in Eqs.~(\ref{h_rf_r_homo1}) and (\ref{h_rf_r_homo2}),
which contain $ \Delta \omega_0 $,
are averaged out, then
Eqs.~(\ref{h_rf_r_homo1}) and (\ref{h_rf_r_homo2}) 
reduce to Eqs.~(\ref{h_rf_r1}) and (\ref{h_rf_r2}).
Simultaneously, we can tune the pulse width $\tau$ short enough
($J \tau \ll 1$)
so that the spin-spin interaction (\ref{h0_homo}) is negligible
while pulses are applied. Therefore we conclude that
an arbitrary pulse sequence designed for 
heteronucleus molecules works
for homonucleus ones provided that
all the hard pulses are replaced by soft pulses
whose pulse width $ \tau $ satisfies the condition
$ 2 \pi/ \left| \Delta \omega_0 \right| < \tau \ll 2 \pi/ J $.
We will demonstrate this consequence experimentally in the next section,
where we set $\tau = 4 (2 \pi/ \left| \Delta \omega_0 \right|) = 5.229$~ms 
$\ll 2 \pi/ J =140.8$~ms.

\subsection{Conventional Hamiltonians}

Here we make comparison between the Hamiltonians derived in the previous 
subsection and the Hamiltonian for homonucleus molecules 
used in literature.
Conventionally the system Hamiltonian
\begin{equation}
\label{h0_homo3}
\tilde{H}_{{\rm conv},0} =
- \Delta \omega_0 I \otimes I_z + J I_z \otimes I_z
\end{equation}
is used to describe
spin dynamics 
in a static magnetic field in a rotating frame \cite{cory2}.
Apparently, it differs from our Hamiltonians (\ref{h0_homo}).

We suspect that
the Hamiltonian (\ref{h0_homo3}) may be derived
from the original system Hamiltonian (\ref{system H})
via transformation from the laboratory frame
to the frame rotating with a common angular velocity
$ \omega_{{\rm rot},1} = \omega_{{\rm rot},2} = \omega_{0,1} $.
We will show, however, that this choice does not yield the Hamiltonian
(\ref{h0_homo3}) in the rotating frame.

If we take a frame that rotates with
a common angular velocity equal to
$\omega_{0,1} $
for both spins,
transformation operator (\ref{rot}) becomes
\begin{equation}
\label{rot1}
U_{\rm com} = 
\exp( - i \omega_{0,1} I_z t) \otimes 
\exp( - i \omega_{0,1} I_z t). 
\end{equation}
Then the system Hamiltonian (\ref{system H}) is transformed into
\begin{eqnarray}
\label{h0_homo2}
\tilde{H}_{{\rm com}, 0} 
&=& 
U_{\rm com} H_0 U_{\rm com}^{\dagger} 
- i U_{\rm com} \frac{d}{dt} U_{\rm com}^{\dagger} 
\nonumber\\
&=&
- \Delta \omega_0 I \otimes I_z 
+ J  I_z  \otimes I_z 
\nonumber \\ && 
+ \left( 
           \begin{array}{cccc}
             0 & 0           &     0       & 0 \\
             0 & 0           & \frac{J}{2} & 0 \\
             0 & \frac{J}{2} &     0       & 0 \\
             0 & 0           & 0           & 0 
           \end{array}
        \right),
\end{eqnarray}
which does not agree with
the conventional Hamiltonian (\ref{h0_homo3}).

Another system Hamiltonian in the laboratory frame
\begin{equation}
\label{h0_homo4}
H_{{\rm conv},0} = 
- \omega_{0,1} I_z \otimes I 
- \omega_{0,2} I \otimes I_z
+ J I_z \otimes I_z
\end{equation}
is also sometimes employed in literature \cite{rmp_chuang,cory,raedt}
but this is also different from
the original system Hamiltonian (\ref{system H}).
We cannot take the Hamiltonian (\ref{h0_homo4}) as a correct one
since we cannot replace 
$ \sum_k J I_k\otimes I_k $ by $ J I_z\otimes I_z $
in the laboratory frame. 

To illustrate the difference between our Hamiltonian 
and the conventional Hamiltonian, let us consider the unitary gate
\begin{eqnarray}
\label{c-phase-gate}
U_{\rm E} 
&=& \exp ( -i \pi I_z \otimes I_z )
\nonumber \\
&=& \left(
\begin{array}{cccc}
e^{-i \pi/4} & 0 & 0 & 0 \\
0 & e^{i \pi/4}  & 0 & 0 \\
0 & 0 & e^{i \pi/4}  & 0 \\
0 & 0 & 0 & e^{-i \pi/4} 
\end{array}
\right),
\end{eqnarray} 
which  is employed along with one-qubit operations
to implement the controlled-NOT
gate \cite{ref:1}. 
We implement the gate $ U_{\rm E} $ from our system Hamiltonian
(\ref{h0_homo}), as
\begin{equation}
\label{UJ}
U_{J} (\pi/J) 
= \exp( -i \pi \tilde{H}_0/J )
= \exp( -i \pi I_z \otimes I_z).
\end{equation}
In other words, we simply wait
for a time interval $T_J = \pi/J $ 
without applying any rf pulses.
Let us define the distance between $ U_{\rm E} $ and $ U_J(t) $ as
\begin{equation}
\| U_{\rm E} - U_J(t) \| 
\equiv 
\sqrt{ {\rm tr}
[ (U_{\rm E} - U_J(t))^{\dagger} (U_{\rm E} - U_J(t)) ] }.
\end{equation}
This is easily evaluated as
\begin{equation}
\| U_{\rm E} - U_{J}(t) \| 
= 
2 \sqrt{2} 
\sqrt{ 1 - \cos \frac{1}{4} (J t - \pi) }.
\label{error1}
\end{equation}
We observe that the distance vanishes at $ t=T_J$ so that
$ U_J(T_J) = U_{\rm E}$. 
We note also that the distance remains close to zero in the vicinity
$t \sim T_J$.
This robust character of $ U_J(t) $ was clearly observed in our experiment
as shown in the next section. 

On the other hand, 
if we replace $\tilde{H}_0$ in Eq. (\ref{UJ}) by the conventional 
Hamiltonian (\ref{h0_homo3}),   
the distance between $ U_{\rm E} $ and $ U_J (t) $ becomes
\begin{eqnarray}
\lefteqn{ \| U_{\rm E} - U_{{\rm conv} J}(t) \| }
\nonumber\\ 
&=& 2 \sqrt{2} 
\sqrt{ 1- 
\cos \Big( \frac{\Delta \omega_0 t}{2} \Big)
\cos \frac{1}{4} ( Jt - \pi )
}.
\label{error2}
\end{eqnarray}
Therefore, if the conventional Hamiltonian (\ref{h0_homo3}) 
were a correct one to describe the spin dynamics,
$ U_{{\rm conv}J} (t) $ would not coincide with $ U_{\rm E} $
at $ t = T_J$
and the distance 
$ \| U_{\rm E} - U_{{\rm conv}J}(t) \| $
should oscillate in the vicinity of the $t \sim T_J$.
However, 
such a rapid oscillation in time has not been observed in our experiment.

\section{Experiments}
\label{exp}

\subsection{Spectrometer and molecules}

All the data were taken at room 
temperature with a JEOL ECA-500 spectrometer \cite{jeol}, where
the hydrogen Larmor frequency is approximately 500~MHz.

We used 0.6 mL, 23 mM sample of cytosine \cite{cytosine}
solved in D$_2$O.
The measured coupling strength is $J/2 \pi=7.1$~Hz 
while the frequency difference is 
$\left| \Delta \omega_0 \right| / 2 \pi = 765.0$~Hz. 
The transverse relaxation time $T_2$ is measured to be 
$\sim 1$~s for both hydrogen nuclei and 
the longitudinal relaxation time $T_1$ is $\sim 7$~s.

In order to measure the spin states, 
we apply a reading pulse to only one spin, called spin 1,
and then obtained the spectrum by Fourier transforming the 
free induction decay (FID) signal. 
The state of the spin 1 is read from the sign of the peak in the spectrum 
while the state of the other nucleus (spin 2) is
found from the peak position.

\subsection{Deutsch-Jozsa algorithm}

The Deutsch-Jozsa (DJ) algorithm \cite{DJ} is one of the simplest 
quantum algorithms 
that illustrate the power of quantum computation 
and has been implemented by several groups \cite{13c,cytosine}.
Let us consider a one-bit function $f: \{0, 1\} \to \{0, 1\}$. 
For a two-qubit register, 
there are only four possibilities for $f$, whose 
explicit forms are
$f_1(0) = f_1(1) = 0$, $f_2(0) = f_2(1)=  1$, $f_3(0) = 0, f_3(1) = 1$
and $f_4(0) = 1, f_4(1) = 0$. The former two functions are said to be
``constant'' while the latter two are ``balanced''.
With the DJ algorithm, we can tell whether a 
given unknown function $f$ is constant or balanced
via only a single trial. 

Chuang {\it et al.} \cite{13c}
employed carbon-13 labeled chloroform, a 
heteronucleus molecule, as a computational resource while
Jones and Mosca \cite{cytosine} used cytosine, 
a homonucleus molecule, to execute the DJ algorithm. 
Chuang {\it et al.} executed the DJ algorithm 
using the pulse sequences shown in Table~\ref{dj_seq}.  
According to our previous discussions, it should be possible to use
the pulse sequences of Chuang {\it et al.} for cytosine molecules
by simply replacing hard pulses with soft ones.

\begin{figure}[b]
\includegraphics[bb=30 0 550 580,width=7.5cm]{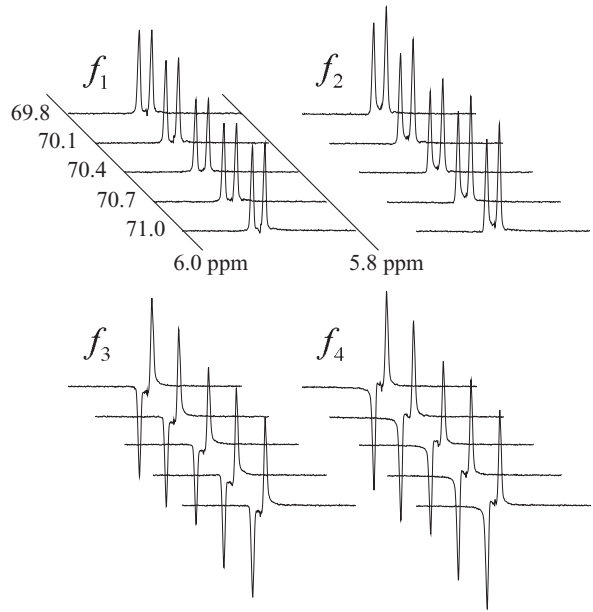}
\caption{
The FID spectra of the spin 1 in cytosine
showing output of the DJ algorithms.
The sign of each peak indicates the state of the spin 1.
The location of each peak indicates the state of the spin 2.
When the initial state of the spin 2 is $|0\rangle$,
it causes a larger shift to the resonance frequency of the spin 1
and contributes to a left peak in each curve. 
Then the sign of the left peak 
discriminates whether $f_i$ is constant or balanced. 
The left peak is
positive for $f_1$ and $f_2$ (constant) and
negative for $f_3$ and $f_4$ (balanced). 
The numbers in the left side 
are durations of the two-qubit operations in ms.
The correct duration of the two-qubit operations 
is $T_J = 70.4$~ms.
Note that the spectra are insensitive to variation
of the two-qubit operation time.
\label{dj_j}}
\end{figure}

The results of our quantum computations with cytosine are 
summarized in Figs. \ref{dj_j} and \ref{dj_p}. 
We started the computation with the thermal equilibrium state 
since the DJ algorithm does not require a
pure initial state~\cite{13c}. 

\begin{table}[t]
\caption{
\label{dj_seq}
Control pulse sequences for the Deutsch-Jozsa algorithm 
taken from the reference \cite{13c}. The functions $f_i(x)$ are 
defined in the text. 
Here, $[\frac{\pi}{2}]_{i}$ denotes
the $\pi/2$ pulses around the $i$-axes $(i=x, y, -x, -y)$.
The symbol $[\pi]$ denotes the $\pi$ pulse 
around the $x$-axis. (1/$nJ$) denotes the two-qubit entangling operation
produced by turning off the rf pulses during the interval $2\pi/nJ$.
The pulse sequences are followed by a readout $\pi/2$ pulse 
around the $x$-axis to the spin 1,
which is not shown in the Table.
The receiver detects the FID (free induction decay) signal 
to measure the spin state.
}
\begin{tabular}{llllllll}
\hline
\hline
\multicolumn{1}{c}{Gate \hspace{3mm} } & 
\multicolumn{7}{c}{Pulse sequence} \\
\hline 
\hspace{2mm} $f_1$ &&&&&&& \\
spin 1 \hspace{4mm}
& $[\frac{\pi}{2}]_y$               &--&(1/4$J$)&  --    &(1/4$J$) &  --    
               &   $[\frac{\pi}{2}]_{-y}$  \\
spin 2 & $[\frac{\pi}{2}]_{-y}$  &--&(1/4$J$)&  $[\pi]$ &(1/4$J$)&  $[\pi]$
               &   $[\frac{\pi}{2}]_y$      \\
\hline 
\hspace{2mm} $f_2$ &&&&&&&\\
spin 1 & $[\frac{\pi}{2}]_y$  &--           &(1/4$J$)  &  --    &(1/4$J$) &  --    
               &  $[\frac{\pi}{2}]_{-y}$  \\
spin 2 & $[\frac{\pi}{2}]_{-y}$  &--&(1/4$J$)& $[\pi]$    &(1/4$J$)& -- 
               &  $[\frac{\pi}{2}]_y$      \\
\hline 
\hspace{2mm} $f_3$  &&&&&&& \\
spin 1 & $[\frac{\pi}{2}]_y$               & -- &  (1/2$J$)  &  $[\frac{\pi}{2}]_{-y}$ 
               & $[\frac{\pi}{2}]_{-x}$& $[\frac{\pi}{2}]_y$  
& $[\frac{\pi}{2}]_{-y}$  \\
spin 2 & $[\frac{\pi}{2}]_{-y}$ & $[\frac{\pi}{2}]_y$   & (1/2$J$) & $[\frac{\pi}{2}]_{-y}$  
               &  $[\frac{\pi}{2}]_{x}$             &-- & $[\frac{\pi}{2}]_y$    \\
\hline 
\hspace{2mm} $f_4$  &&&&&&&\\
spin 1 & $[\frac{\pi}{2}]_y$               &-- &  (1/2$J$)  & $[\frac{\pi}{2}]_{-y}$ 
               & $[\frac{\pi}{2}]_{-x}$ &$[\frac{\pi}{2}]_y$ & $[\frac{\pi}{2}]_{-y}$  \\
spin 2 & $[\frac{\pi}{2}]_{-y}$  & $[\frac{\pi}{2}]_y$   & (1/2$J$) & $[\frac{\pi}{2}]_{-y}$  
               & $[\frac{\pi}{2}]_{-x}$ & -- & $[\frac{\pi}{2}]_y$           \\
\hline 
\hline
\end{tabular}
\end{table}

\subsubsection{$J$-coupling time\label{J-c}}

The DJ algorithm employs the $ J $-coupling unitary operator 
$ U_J(t) $
with 
$ t = 2 \pi / 4J = T_J/2$ or $ t = 2 \pi / 2J =T_J$
to entangle two spins.
The time durations for the two-qubit operations are
$ 2 \pi/4J + 2 \pi/4J = T_J $ for $f_1$ and $f_2$ and
$T_J$ for $f_3$ and $f_4$.
Thus the total execution time is $T_J$ for all four cases.

As we discussed when we derived Eq. (\ref{error1}),
our Hamiltonian (\ref{h0_homo}) predicts that 
$ U_J(t) $ does not deviate much from the desired unitary transformation
even when the gate operation time $ t $ deviates 
from the correct value.
On the other hand, Eq. (\ref{error2}) tells us that
the conventional Hamiltonian (\ref{h0_homo3}) predicts that
$ U_J(t) $ sharply depends on the timing $ t $ and oscillates as
$ \cos (\Delta \omega_0 t/2) $.

In experiment we executed the DJ algorithm 
with various gate operation time $t$ in the vicinity of $T_J$
and observed how the resulting spectra depend 
on $t$.
We employed the pulse sequences shown in Table~\ref{dj_seq},
in which all hard pulses used by Chuang {\it et al.} \cite{13c} were 
replaced with Gaussian soft pulses with the pulse width 5.229~ms.

The initial state of the molecules
is a thermal mixture of four states $|00\rangle$, $|01\rangle$, 
$|10\rangle$, and $|11\rangle$. 
The DJ algorithm does not work when the second qubit is $|1\rangle$
and fails to distinguish constant from balanced. 
On the other hand, 
it works regardless of the state of the first qubit. 
In Fig.~\ref{dj_j} 
the peaks with a smaller frequency shift (the right peaks) 
are outputs from the initial states 
$ | 01 \rangle $ or
$ | 11 \rangle $. 
In this case the algorithm fails
to distinguish if $f_i$ is constant or balanced.
The peaks with a larger frequency shift (the left peaks) 
in Fig.~\ref{dj_j} 
are outputs from the initial states
$ | 00 \rangle $ or
$ | 10 \rangle $. 
In this case the DJ algorithm successfully
tells us if $f_i$ is constant or balanced
by the sign of the peak (positive for $f_1$ and $f_2$ while negative for $f_3$
and $f_4$).

We varied the $ J $-coupling time interval $ t $
in the range from 69.8~ms to 71.0~ms.
In other words,
$ \Delta \omega_0 t /2 $
was swept between 
$ 26.7 \times 2 \pi$ and $ 27.2 \times 2 \pi $. 
The exact duration to produce the designed unitary operator 
correctly is 70.4~ms.
We observe from Fig.~\ref{dj_j} that
the spectra are not sensitive to variation of the time interval.
Therefore we concluded that our Hamiltonian (\ref{h0_homo}) 
accounts for the experimental results consistently.

\subsubsection{Rf pulse width}

In literature \cite{cytosine, raedt} it is recommended to use
soft pulses whose width is an integral multiple
of $ 2 \pi / \left| \Delta \omega_0 \right| $ 
in order to avoid undesirable effect caused by the term 
$ \Delta \omega_0 I \otimes I_z $ 
in the conventional Hamiltonian (\ref{h0_homo3}). 
Our discussion and experiment show that
this tuning is not necessary since
the relevant Hamiltonian (\ref{h0_homo}) does not contain
the term $ \Delta \omega_0 I \otimes I_z $.

\begin{figure}[b]
\includegraphics[bb=50 50 450 450,width=7.5cm]{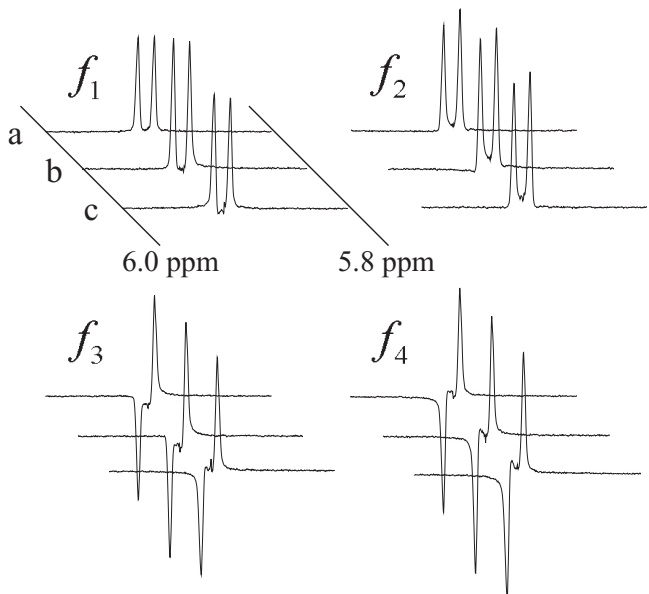}
\caption{
The effect of variation of the pulse widths on the spectrum of the DJ algorithms. 
The pulse widths for spins 1 and 2 are 
(a) 5.229~ms and 5.229~ms, 
(b) 6.217~ms and 6.217~ms, and
(c) 5.229~ms and 6.217~ms, respectively. 
Observe that the spectra are insensitive to the variation of pulse width.
\label{dj_p}}
\end{figure}

We executed the DJ algorithms shown in Table~\ref{dj_seq}, 
with different pulse width (5.229 and 6.217~ms).
In our setting,
the pulse width 5.229~ms is equal to 
$ 4 \times 2 \pi / \left| \Delta \omega_0 \right|$,
while 6.217~ms is $ 4.76 \times 2 \pi / \left| \Delta \omega_0 \right|$.
The measured FID spectra of the spin 1 are shown in Fig.~\ref{dj_p}.
No significant changes in the spectra appeared 
even if we tuned the pulse width to a fractional multiple of 
$ 2 \pi/ \left| \Delta \omega_0 \right|$. 
This result proves that
the pulse width need not be 
an integral multiple of $ 2 \pi / \left| \Delta \omega_0 \right| $
to implement a given gate and obtain a reasonable spectrum.

\section{Field Inhomogeneity Compensation}

Here we discuss an experiment
to reveal the nature of the Hamiltonians
(\ref{h_rf_r_homo1}) and
(\ref{h_rf_r_homo2}),
which depict the action of the oscillating magnetic fields 
on the spins.
It is common to employ the compensating pulse method to
suppress errors induced by field inhomogeneity.
We will show, by employing our Hamiltonian, that
the entangling operation with the $J$-coupling
is fragile in the presence of 
the compensating pulses and a fine tuning of the gate operation time is
required.

\subsection{$\pi$-pulse pair in J-coupling time}

We have shown in the previous section that 
it is not necessary to tune the $ J $-coupling time very accurately
since this operation is robust against small change of the
gate operation time.
However, if the system is under the influence of field inhomogeneity,
it may cause an error during the $ J $-coupling time.

It is well known that this undesired effect caused by 
field inhomogeneity 
can be compensated by a series of hard 
$\pi$-pulse pair, whose width is of the order of 10~$\mu$s.
The best known example may be
the CPMG (Carr-Purcell-Meiboom-Gill) pulse sequence~\cite{NMR_textbook}.  
Let us apply this technique to $U_{J} (t) $.
Then the pulse sequence for $ U_J(t) $ is replaced with 
\begin{equation}
U_{J*}(t) \, : \, U_J(t/2)-[\pi]-U_J(t/2)-[\pi], 
\label{J star pulse}
\end{equation}
where time flows from left to right and
$[\pi]$ denotes a hard pulse 
that rotates both spins by $\pi$ radian. We take the rotation axis
to be
the $x$-axis of one of the spins, for example the spin 1, while
the rotation axis for the other spin, the spin 2, depends on the time when 
the $[\pi]$ is applied, according to the Hamiltonian (\ref{h_rf_r_homo1}).
The pulse sequence is represented as the product of unitary matrices
\begin{equation}
U_{J*}(t) = 
U_{1,\pi}^{(2)} \, U_J(t/2) \, U_{1,\pi}^{(1)} \, U_J(t/2) ,
\end{equation}
where
\begin{eqnarray}
U_{1,\pi}^{(1)}
& = &
\exp ( -i \pi I_x \otimes I )
\\ \nonumber 
& \times & 
\exp (
-i \pi [ 
  \cos (\Delta \omega_0 \frac{t}{2}) I \otimes I_x
+ \sin (\Delta \omega_0 \frac{t}{2}) I \otimes I_y
] ) , \\
U_{1,\pi}^{(2)}
& = &
\exp ( -i \pi I_x \otimes I )
\\ \nonumber 
& \times & 
\exp \left(
-i \pi [ 
  \cos (\Delta \omega_0 t) I \otimes I_x
+ \sin (\Delta \omega_0 t) I \otimes I_y
] \right) . 
\end{eqnarray}
Note that 
we put the ratio of resonance frequencies
$ g = \omega_{0,2} / \omega_{0,1} = 1 $
for the homonucleus molecule.
The resulting operator $ U_{J*} (t) $ 
does not coincides with $ U_J (t) $.
The distance between $ U_{J*}(t) $ and $ U_{\rm E}$ is evaluated as
\begin{eqnarray}
\label{error3}
\lefteqn{
\| U_{\rm E} - U_{J*}(t) \| 
}
\nonumber\\  
&=& 2 \sqrt{2}
\sqrt{1-
\cos \Big( \frac{\Delta \omega_0 t}{2} \Big)
        \cos \frac{1}{4} (Jt - \pi) 
}.
\end{eqnarray}
The distance does not vanish generally
because the two conditions,
$ \cos ( \Delta \omega_0 t / 2 ) = \pm 1 $ and
$ \cos ( J t/4 - \pi/4 ) = \pm 1 $,
are rarely satisfied simultaneously. 
Moreover, the distance is very sensitive to $t$.

\begin{table}[b]
\caption{
\label{tb_pps}
Control pulse sequences to create the pseudo-pure state $|00\rangle$
\cite{pps_j}.
Here, $[\frac{\pi}{m}]_{i}$ denotes
the $ \frac{\pi}{m} $ pulse around the $i$-axes $(i=x, y, -x, -y)$
while
(1/2$J$) denotes the two-qubit entangling operation
implemented by turning off the rf pulses during the interval $2\pi/2J$.
The symbol
FG denotes application of a pulsed field gradient for spatial labeling. 
The pulse sequence is followed 
by a readout $\pi/2$-pulse around the $x$-axis to the spin 1,
which is not shown in the Table.
}
\begin{tabular}{lcccccc}
\hline
\hline
\multicolumn{1}{c}{} & 
\multicolumn{6}{c}{Pulse sequence} \\
\hline 
spin 1 \hspace{3mm} 
&               & FG  & $[\frac{\pi}{4}]_x $&(1/2$J$) &$[\frac{\pi}{4}]_{-y}$&FG \\
spin 2 
& $[\frac{\pi}{3}]_x$ & FG &                &(1/2$J$) &                &   FG  \\
\hline
\hline 
\end{tabular}
\end{table}

\subsection{Pseudo-pure state preparation}

\begin{figure}[t]
\includegraphics[bb=0 0 585 250,width=7.5cm]{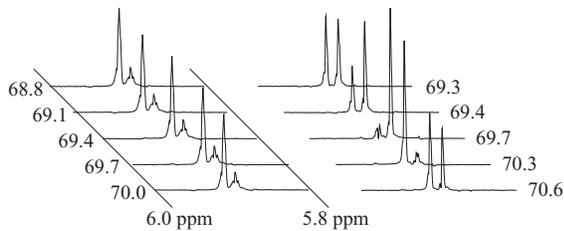}
\caption{
The effect of variation of the $J$-coupling time on the spectrum of the 
pseudo-pure state creation. 
The spectra in the left panel are measured
without the $\pi$-pulse pair in the entangling operation.
The spectra in the right panel are measured
with the $\pi$-pulse pair.
The numbers are the two-qubit operation time in ms.
Note that the spectra
generated with the $\pi$-pulse pair are 
sensitive to variation of the two-qubit operation time.
\label{fig:pps}}
\end{figure}

An NMR quantum computer must be initialized to the pseudo-pure state 
$ |00 \rangle$ before executing a specific algorithm.
This initialization procedure is implemented with
the pulse sequence~\cite{pps_j} shown in Table~\ref{tb_pps}.
In this section we examine the effect of the compensating $ \pi $ pulses
on the initialization procedure.

The initialization process contains the $ J $-coupling time,
in which the two spins are entangled by the two-qubit operation  $ U_J(t) $.
We vary the gate operation time
$t$ to introduce an operational error on purpose.
It is possible, however, to apply a pair of $ \pi $-pulses to 
compensate this error while the $J$-coupling is under action as 
instructed in Eq. (\ref{J star pulse}).
We have swept the gate operation time between 69.3~ms and 70.6~ms and  
the results are summarized in Fig.~\ref{fig:pps}.

The spectra in the left panel of Fig.~\ref{fig:pps}
were measured {\it without applying} the $ \pi $-pulse pair
in the entangling operation.
We observed that the spectra are robust
against small variations of the gate operation time.
The intense peaks with a larger frequency shift
are signals from molecules in the $ |00\rangle $ state while
the smaller peaks are error signals from
small amount of molecules not in the $ |00 \rangle $ state. 

The spectra in the right panel 
were measured with the $\pi $-pulse pair applied
during the entangling operation.
We observed that the spectra are very sensitive to the variation 
of the $ J $-coupling operation time,
although it is still possible to adjust the operation time so that the 
desired pseudo-pure state is produced with a good precision.
When the gate operation time was set at the correct value $ t = 70.3$~ms, 
the spectrum exhibited a sharper peak than those in the left panel.
This result implies 
that the $ \pi $-pulse pair improved the quality of the initialized state.
However, the spectra were fragile
when the duration $t$ deviates from the correct value.
For example, when the duration was set at $ t=69.7$~ms,
the intense signal indicated that
the most of molecules are in the $ |01 \rangle $ state, and
not in the desired state $|00\rangle $. 

Thus we conclude that
our Hamiltonian (\ref{h_rf_r_homo1}) and (\ref{h_rf_r_homo2})
accurately account for the experimental results.

\vspace{4ex}
\section{Conclusions and Discussion}

We have derived the relevant Hamiltonian for homonucleus molecules
in NMR quantum computing and shown that any pulse sequence for
a heteronucleus molecule may be translated into that for a homonucleus
molecule by simply replacing hard pulses by soft pulses with
a properly chosen pulse width.
We have demonstrated that the NMR spectra in several experiments are
accounted for with our Hamiltonian but not with the conventional
Hamiltonian found in literature. It was shown in our experiments that
the spectra are robust
under small variations of the $J$-coupling operation time
as well as of the rf pulse widths.
Moreover, we provided the theoretical basis for
field inhomogeneity compensation by a pair of hard $\pi$ pulses 
during the entangling operation and verified it experimentally.

Generalization of the present work to molecules with more spins
is straightforward. It is easy to find proper pulse sequence, either
numerically \cite{qaa} or by Cartan decomposition \cite{ref:kg, warp}, 
once an exact form of the
Hamiltonian is obtained. Theoretical analysis as well as experiments
on these subjects are under progress and will be published elsewhere.

\section*{Acknowledgments}

We would like to thank Manabu Ishifune for sample preparation,
Toshie Minematsu for assistance in
NMR operations and Katsuo Asakura and Naoyuki Fujii of JEOL
for assistance in NMR pulse programming.
MN would like to thank partial supports of Grant-in-Aids for 
Scientific Research from the Ministry of Education, 
Culture, Sports, Science and Technology, Japan, Grant No.~13135215 and
from Japan Society for the Promotion of Science, Grant No.~14540346.
ST is partially supported by the Ministry of Education, Grant No.~15540277.

\end{document}